# Resonances in the system of $\pi^+\pi^-$-mesons from the reaction $np \to np\pi^+\pi^-$ at $P_n = 5.20\, GeV/c$: search, results of direct observations, interpretation.

Yu.A. Troyan[1][†], E.B. Plekhanov[1], V.N. Pechenov[1], A.Yu. Troyan[1],
A.V. Beljaev[1], A.P. Jerusalimov[2], S.G. Arakelian[1]

[1]-Veksler and Baldin Laboratory of High Energies, JINR.
[2]- Laboratory of Information Technologies, JINR

[†]-troyan@sunhe.jinr.ru

10 resonances were found in the mass spectrum of $\pi^+\pi^-$-system based on 66075 events from the reaction $np \to np\pi^+\pi^-$ in $np$-interactions at $P_n = (5.20 \pm 0.16)\, GeV/c$ in the 1-m HBC of LHE JINR by using the criterion $\cos\Theta^*_p > 0$. These masses are the following: (347 ± 12), (418 ± 6), (511 ± 12), (610 ± 5), (678 ± 17), (757 ± 5), (880 ± 12), (987 ± 12), (1133 ± 15), and (1285 ± 22) MeV/$c^2$; their excess above background is 2.9, 5.2, 3.5, 1.4, 2.0, 8.5, 4.8, 3.8, 5.2, and 6.0 S.D., respectively. The experimental widths of the resonances vary within the region from 16 to 94 MeV/$c^2$. Such effects were not found in $\pi^-\pi^0$-combinations from the reaction $np \to pp\pi^-\pi^0$. Therefore, it is necessary to attribute the value of isotopic spin I = 0 to the resonances found in the mass spectrum of the $\pi^+\pi^-$-system. The spin was estimated for the most statistically provided resonances at masses of 418, 511 and 757 MeV/$c^2$. We determine with a high degree of confidence that J = 0 for the resonances at $M_R = 757$ MeV/$c^2$ and $M_R = 418$ MeV/$c^2$ and the most probable value of J = 0 for the resonance at $M_R = 511$ MeV/$c^2$.

**Therefore, it can be affirmed that at least 3 states with quantum numbers of $\sigma_0$-meson $0^+(0^{++})$ have been found at masses of 418, 511 and 757 MeV/$c^2$.**

The fact low-mass $\sigma_0$-mesons are glueballs is one of the possible interpretations.
The comparison with the data of other papers has also been made.

The investigation has been performed at the Veksler and Baldin Laboratory of High Energies, JINR.

# 1. Introduction

This work is devoted to a search for and study of low-mass ($M < 1.3\ GeV/c^2$) resonances in the $\pi^+\pi^-$-system. Their existence can clarify the properties of low-lying scalar mesons (the so-called $\sigma_0$-mesons), whose investigation is important both for the mechanism of realization of chiral symmetry for corresponding Lagrangians and for an adequate description of an attractive part of the nucleon-nucleon interaction potential [1].

Different theoretical models give various predictions for masses and widths of $\sigma_0$-mesons. Early quark bag models gave $M_{\sigma_0} > 1.5\ GeV/c^2$ and $\Gamma_{\sigma_0} \approx 0.5\ GeV/c^2$ [2]. Later works predicted $M_{\sigma_0} = 500 \div 1000\ MeV/c^2$ and $\Gamma_{\sigma_0} = 200 \div 500\ MeV/c^2$ for low-lying $(q\bar{q})$-states [3]. Some models of a spontaneous break of chiral symmetry predict $M_{\sigma_0} \approx 700\ MeV/c^2$ and $\Gamma_{\sigma_0} \approx 500\ MeV/c^2$ [4]. Using QCD sum rules and assuming that the $\sigma_0$-meson is a low-lying glueball, the calculations give the following predictions: $M_{\sigma_0} = 280 \div 700\ MeV/c^2$ and $\Gamma_{\sigma_0} = 2 \div 60\ MeV/c^2$ [5] (see also [6]).

# 2. Selection of kinematics criteria and results of investigations

This paper continues a series of works, devoted to the study of the $\pi^+\pi^-$-system, with different kinematics criteria [7,8].

66075 events of the reaction $np \to np\pi^+\pi^-$ at $P_n = 5.20\ GeV/c$ have been treated. The data were obtained in an exposure of the 1m $H_2$ bubble chamber of LHE (JINR) to a monochromatic neutron beam ($\Delta P_n / P_n \approx 2.5\%$, $\Delta\Omega_{channel} = 10^{-7}\ sterad.$) [9].

The events of various reaction channels were separated by the standard $\chi^2$–methodic [10-12].

Fig.1 shows the effective mass distribution of $\pi^+\pi^-$-combinations from the total statistics of the reaction $np \to np\pi^+\pi^-$ at $P_n = 5.20\ GeV/c$. The distribution is approximated by a polynomial background curve and by 3 resonance curves taken in the Breit-Wigner form. 3 resonance peaks are found at the masses of 418, 511 and 757 $MeV/c^2$. The excess is more than 3 S.D. above background.

Fig.2 shows the effective mass distribution of $\pi^+\pi^-$-combinations for the events with a secondary neutron flying in the forward hemisphere in c.m.s. of the reaction, i.e. $\cos\Theta_n^* > 0$. No noticeable deviations above background have been observed in this distribution.

Earlier, we have already studied the reaction $np \to np\pi^+\pi^-$ [13], and OPE-exchange with a dominated exchange of the charged $\pi$-meson has been shown to be main mechanism of this reaction. It leads to a plentiful production (up to 70% of the total reaction cross section) of $\Delta^{++}$ and $\Delta^-$-resonances in the lower and upper vertices of the corresponding diagrams. The OPE mechanism gives a main part into the events with neutron flying into the forward hemisphere.

Therefore, it seems reasonable to study the resonances in the $\pi^+\pi^-$-system of the reaction $np \to np\pi^+\pi^-$ selecting the events on condition that $\cos\Theta_p^* > 0$. The total contribution of the $\Delta^{++}$ and $\Delta^-$-resonances is no more than 17% for these events, and the background from resonance decays decreases greatly. The number of events with $\cos\Theta_p^* > 0$ is equal to 20266, that is approximately 1/3 of all the events from the reaction $np \to np\pi^+\pi^-$.

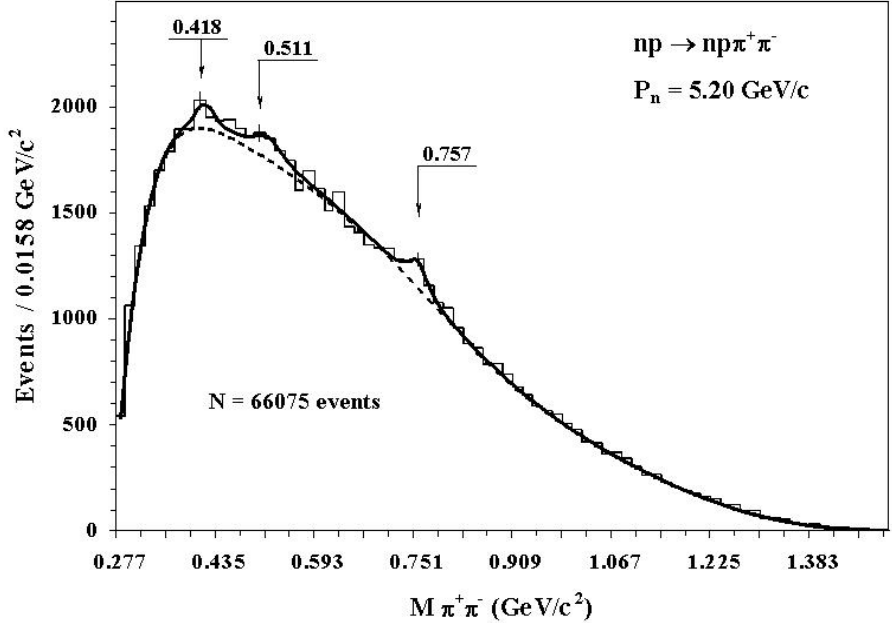

**Fig. 1**

The effective mass distribution of $p^+p^-$-combinations from the total statistics of the reaction $np \to npp^+p^-$ at $P_n = 5.20\,GeV/c$.

The dotted curve is the background taken in the form of a superposition of Legendre polynomials up to the 10-th degree, inclusive.

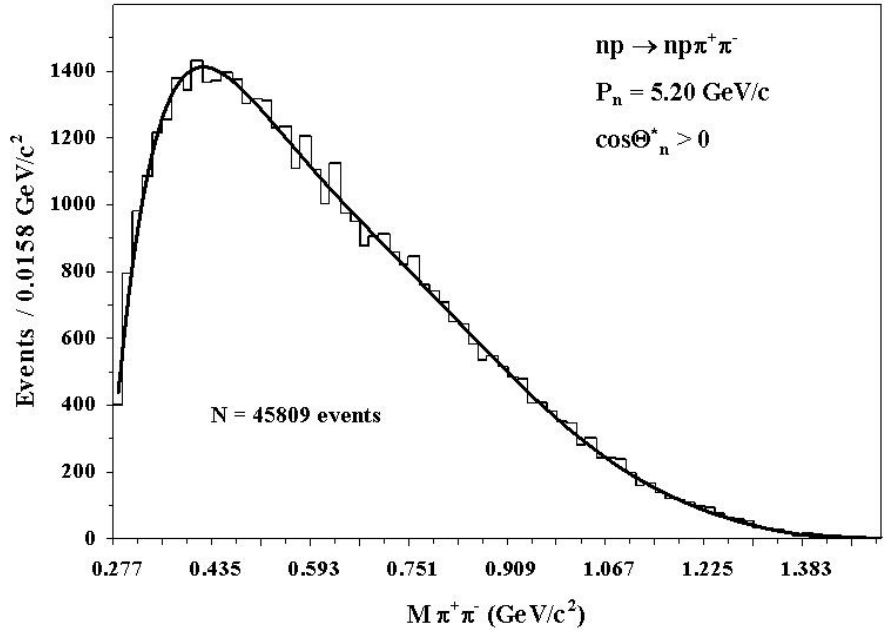

**Fig. 2**

The effective mass distribution of $p^+p^-$-combinations from the reaction $np \to npp^+p^-$ at $P_n = 5.20\,GeV/c$, selected under condition of $\cos\Theta_n^* > 0$.

The solid curve is the background taken in the form of a superposition of Legendre polynomials up to the 8-th degree, inclusive.

Fig.3 shows the effective mass distribution of $p^+p^-$-combinations for the events with a secondary proton flying into the forward hemisphere in the c.m.s. of the reaction.

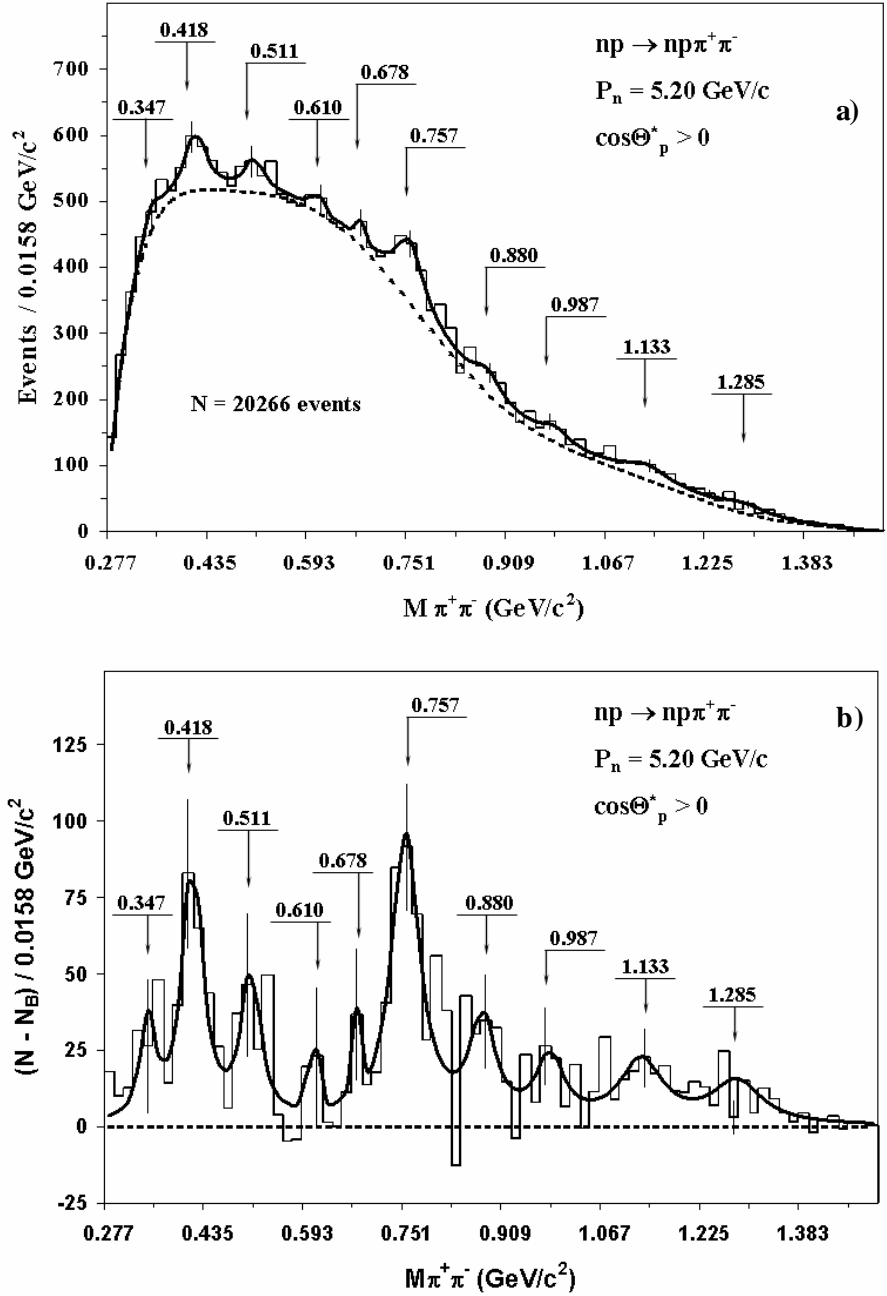

**Fig. 3**

**a)** The effective mass distribution of $p^+p^-$-combinations from the reaction $np \to np p^+ p^-$ at $P_n = 5.20\,GeV/c$, selected under condition of $\cos\Theta_p^* > 0$.

The dotted curve is the background taken in the form of a superposition of Legendre polynomials up to the 9-th degree, inclusive.

The solid curve is the sum of the background and 10 Breit-Wigner form resonance curves.

**b)** The same as fig. 3a, minus background

The distribution is approximated by a polynomial background curve and by 10 resonance curves taken in the Breit-Wigner form (Fig.3 a). The description of the intervals between the resonances by the background gives $c^2 = 0.98 \pm 0.19$ and $\sqrt{D} = 1.47 \pm 0.14$. This is close to theoretical values of 1 and 1.41 for the distribution of random values at 1 constraint equation (normalization to the number of events). The description of the full distribution by the same background, normalized to 100% experimental events gives $c^2 = 1.28 \pm 0.11$ and $\sqrt{D} = 1.60 \pm 0.14$. The description of the full distribution by background and 10 resonance curves gives $c^2 = 0.81 \pm 0.11$ and $\sqrt{D} = 1.33 \pm 0.08$. The same distribution minus background is shown in Fig. 3 b.

The results of approximation are given in the Table I.

**Table I**

|    | $M_e \pm \Delta M_e$ $MeV/c^2$ | $\Gamma_e \pm \Delta\Gamma_e$ $MeV/c^2$ | $\sigma_{mb}$ | S.D. |
|----|---|---|---|---|
| 1  | 347 ± 12  | 36 ± 35 | 10 ± 5 | 2.9 |
| 2  | 418 ± 6   | 39 ± 13 | 26 ± 7 | 5.2 |
| 3  | 511 ± 12  | 40 ± 23 | 15 ± 6 | 3.5 |
| 4  | 610 ± 5   | 24 + 13 | 5 ± 5  | 1.4 |
| 5  | 678 ± 17  | 16 + 14 | 6 ± 4  | 2.0 |
| 6  | 757 ± 5   | 51 ± 15 | 38 ± 7 | 8.5 |
| 7  | 880 ± 12  | 45 ± 24 | 14 ± 5 | 4.8 |
| 8  | 987 ± 12  | 49 ± 36 | 11 ± 4 | 3.8 |
| 9  | 1133 ± 15 | 80 ± 30 | 10 ± 3 | 5.2 |
| 10 | 1285 ± 22 | 94 ± 30 | 10 ± 2 | 6.0 |

*The first column* contains the experimental values of the resonance masses (including errors) obtained in the process of approximation.

*The second column* contains the experimental values of the resonance widths.

*The third column* contains the production cross sections for the corresponding resonances. For the in the cross sections errors, we have taken into account the cross section error for the reaction $np \to np\pi^+\pi^-$ at $P_n = 5.20\, GeV/c$ ($\sigma_{np \to np\pi^+\pi^-} = (6.22 \pm 0.28)\, mb$) [11].

*The fourth column* contains the number of standard deviations of the effects above background: $S.D. = N_{res.} / \sqrt{N_{back.}}$.

The observed resonance at the mass of $M_R = 757\, MeV/c^2$ has been already inserted in RPP-2000, RPP-2002 (S.D.=6.0) [16]. It should be noted that the increase of S.D. in comparison with the previously obtained results corresponds exactly to the increase of the event statistics.

### 3. Spin and isotopic spin of the resonances

To determine the spin of the resonances, the angular distributions of $\pi$-mesons were studied in the helicity coordinate system [14].

For strong decays, such distributions should be described by the sum of Legendre polynomials of even degree with a maximum power of $2J$, where $J$ is the resonance spin.

The distributions of this angle are shown in Fig. 4. for the resonances at $M_R = 418$, 511 and $757\, MeV/c^2$.

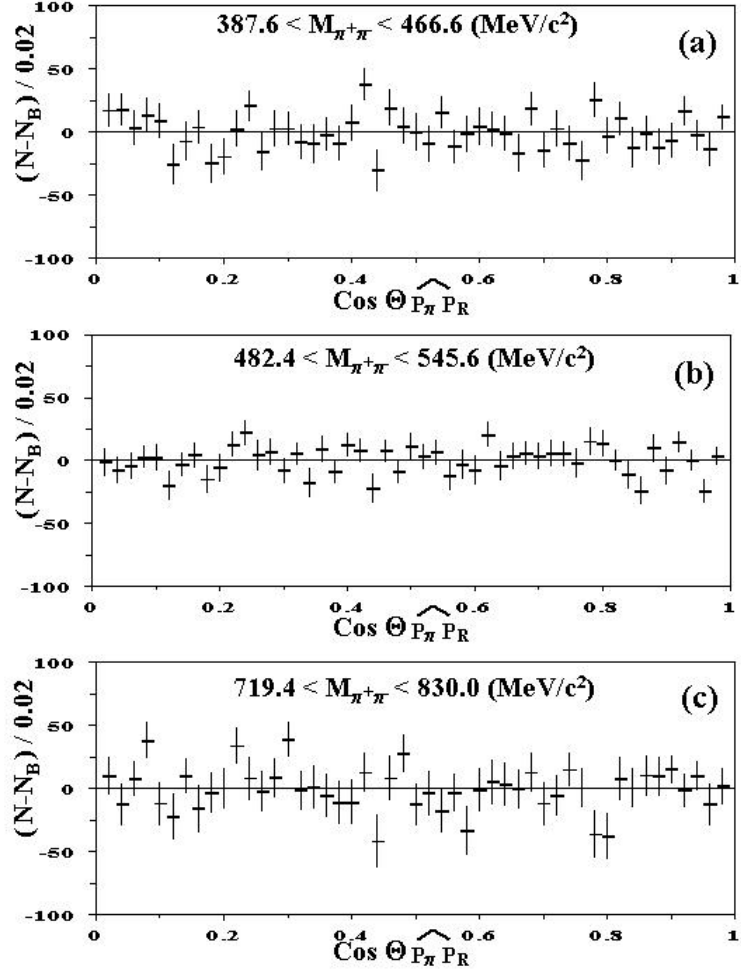

**Fig. 4**

The distribution of decay angle for $p^+p^-$-resonances

**a)**     for the resonance with $M_R = 418\ MeV/c^2$,

**b)**     for the resonance with $M_R = 511\ MeV/c^2$,

**c)**     for the resonance with $M_R = 757\ MeV/c^2$,

The solid lines correspond to the isotropic distribution

The solid lines correspond to the isotropic distribution $(J = 0)$. Therefore, the most probable spin values for the resonances at masses of 418, 511 and 757 $MeV/c^2$ are equal to 0.

The spins of other resonances have not been determined due to low statistics.

From the generalized Pauli principle for the $2p$-system, it follows that the isotopic spin at $J = 0$ must be only even. As it follows from isotopic relations for $I = 2$, the effects at the corresponding masses in the $p^-p^0$-system from the reaction $np \to pp p^- p^0$, that has also been studied by us, could be observed with a statistical significance by 4.5 times more than in the reaction $np \to np p^+ p^-$. But there are no such effects in the $p^-p^0$-system [7].

**Therefore, it can be affirmed that at least 3 states have the quantum numbers $I^G(J^{PC}) = 0^+(0^{++})$ and may be identified as $s_0$-mesons at the masses of 418, 511 and 757 MeV/$c^2$.**

## 4. Interpretation of low-masses $\sigma_0$-mesons as glueballs

The fact low-mass $\sigma_0$-mesons are glueballs is one of the possible interpretations.

The estimate of the width of the $\sigma_0$-glueball given in paper [6] is based on the low energy theorems: $\Gamma_{(\sigma_0 \to \pi\pi)} = \left(M/1GeV\right)^5 \begin{Bmatrix} 220 \\ 550 \end{Bmatrix} MeV/c^2$, where 220 and 550 are the values from a gluon condensate corresponding to 2 variants of the theory.

Let us consider the resonance at of $M_R = 757\ MeV/c^2$. In our experiment, the mass resolution for the $\pi^+\pi^-$–system depends on mass and equals to $\Gamma_{resol}(M) = 0.042(M - 2m_\pi) + 2.8$ (all the values are in $MeV/c^2$) [15]. The width of the resolution function is $\Gamma_{resol.}(757) = 23\ MeV/c^2$ for $M_R = 757\ MeV/c^2$. If the mass resolution function has a normal distribution form (realized for $M_R = 757\ MeV/c^2$), the true width of the resonance can be estimated equal to: $\Gamma_R = \sqrt{\Gamma_{exp.}^2 - \Gamma_{resol}^2} = \sqrt{(51 \pm 15)^2 - (23)^2} = 46^{+16}_{-18}\ MeV/c^2$ and the true width of the resonance is within the range from 36 to $66\ MeV/c^2$.

If the resonance at a mass of $757\ MeV/c^2$ is a glueball, then its width determined by the formula from the low-energy theorems is: $\Gamma_{(\sigma \to \pi\pi)} = \left[(0.757)^5\right]^{220}_{550} = [0.249]^{220}_{550} = ^{55}_{137}\ MeV/c^2$

Therefore, the width of the $\sigma_0$-meson at a mass of $757\ MeV/c^2$ determined in our experiment does not contradict its interpretation as a glueball.

The widths of the resonances at masses of 418 and $511\ MeV/c^2$ do not contradict their interpretation as glueballs either.

## 5. Comparison with other data

A large number of publications are dedicated to the search and study of $\sigma_0$-mesons (see [16]). All of them are based on the PWA of $\pi N$ or $\tilde{p}p$ - interactions. The obtained $\sigma_0$-meson masses ranging from 400 to $1200\ MeV/c^2$ coincide with the mass sequence observed in our experiment. We should emphasize that our results of this study are based on the observed direct signals from the resonances in the effective mass spectra of the corresponding particle combinations unlike all the works devoted to the search for low-mass ($M < 1.3\ GeV/c^2$) $\sigma_0$-mesons. However, the resonance widths extracted from PWA are considerably larger than those obtained in our experiment. It may be necessary to use other ideas and more complicated methods of analysis to understanding these phenomena better.

## 6. Acknowledgement

We are grateful to Dr. V.L.Lyuboshits for his help in our work.